\documentclass[twocolumn]{jpsj3}
\usepackage{times}
\usepackage[dvips]{color}

\def\ef{\epsilon_{\rm F}}

\def\lsim{\lower -0.3ex \hbox{$<$} \kern -0.75em \lower 0.7ex \hbox{$\sim$}}
\def\gsim{\mathrel{\rlap{\lower4pt\hbox{\hskip1pt$\sim$}}
   \raise1pt\hbox{$>$}}}                

\def\tn{$T_{\rm AF}$}

\def\jo #1#2#3#4{#1 {\bf #2}, #4 (#3)}

\def\PRB{Phys.\ Rev.\ B}
\def\PRL{Phys.\ Rev.\ Lett.}
\def\JCP{J.\ Chem.\ Phys.}

\def\JPSJ{J.\ Phys.\ Soc.\ Jpn.}

\def\SSC{Solid\ State\ Commun.}
\def\SM{Synth.\ Met.}
\def\BCSJ{Bull.\ Chem.\ Soc.\ Jpn.}

\def\JPF{J.\ Phys.\ France}
\def\JPCM{J.\ Phys.:\ Condens.\ Matter}
\def\CR{Chem.\ Rev.}

\def\JACS{J.\ Am.\ Chem.\ Soc.}

\def\CPL{Chem.\ Phys.\ Lett.}
\def\MCLC{Mol. Cryst. Liq. Cryst.}

\usepackage{bm} 

\definecolor{green}{rgb}{0.0, 0.5, 0.0}

\usepackage{ulem}

\title{
Fragment Model Study of Molecular Multiorbital System 
$X$[Pd(dmit)$_2$]$_2$
}

\author{%
Hitoshi~Seo$^{1,2,3}$\thanks{E-mail: seo@riken.jp},
Takao~Tsumuraya$^{4,5}$, 
Masahisa~Tsuchiizu$^{6}$, 
Tsuyoshi~Miyazaki$^{5}$, 
and Reizo~Kato$^{4}$
}

\inst{%
$^1$Condensed Matter Theory Laboratory, RIKEN, Wako, Saitama 351-0198, Japan\\
$^2$ RIKEN Center for Emergence Matter Science (CEMS), Wako, Saitama 351-0198, Japan\\
$^3$JST, CREST, Wako, Saitama 351-0198, Japan\\
$^4$Condensed Molecular Materials Laboratory, RIKEN, Wako, Saitama 351-0198, Japan\\
$^5$National Institute for Materials Science, Tsukuba, Ibaraki 305-0047, Japan\\
$^6$Department of Physics, Nagoya University, Nagoya 464-8602, Japan\\
}

\abst{
The electronic properties of quasi-two-dimensional molecular conductors 
 $X$[Pd(dmit)$_2$]$_2$ are studied theoretically. 
We construct an effective model based on the fragment molecular orbital scheme 
 developed recently, 
 which can describe the multiorbital degree of freedom in this system. 
The tight-binding parameters for a series of $\beta'$-type compounds with different cations $X$ 
 are evaluated by fitting to 
 first-principles band calculations. 
We find that the transfer integrals within the dimers of Pd(dmit)$_2$ molecules, 
 along the intramolecular and intermolecular bonds including the diagonal ones, 
 are of the same order, 
 leading to hybridization between different molecular orbitals. 
This results in charge disproportionation within each molecule, 
 as shown in our previous ab initio study 
 [Tsumuraya et al., J. Phys. Soc. Jpn. {\bf 82}, 033709 (2013)], 
 and also leads to a revised picture of the effective dimer model. 
Furthermore, we discuss broken-symmetry insulating states 
 triggered by interaction effects, 
 which show characteristic features owing to the multiorbital nature. 
The on-site Coulomb interaction induces antiferromagnetic states with 
 an intramolecular antiparallel spin pattern, 
 while electron-lattice couplings stabilize nonmagnetic charge-lattice-ordered states 
 where two types of dimer with different charge occupations are arranged periodically. 
These states showing different spatial patterns 
 compete with each other as well as with 
 the paramagnetic metallic state. 
}

\begin{document}
\maketitle

\section{Introduction} 
\label{sec1}

Quasi-two-dimensional molecular conductors
 $X$[Pd(dmit)$_2$]$_2$ (dmit = 1,3-dithiole-2-thione-4,5-dithiolate)
 show various properties 
 depending on the choice of $X$ 
 and the application of pressure~\cite{Kato_2004CR,Tamura_2009STAM,Kato_2014BCSJ}. 
Among them, 
 a series of isostructural $\beta'$-type compounds have been systematically investigated: 
 $X$ = Et$_{4-y}$Me$_yZ$ with $y$ = 0 -- 2, 
 Et = C$_2$H$_5$, Me = CH$_3$, and $Z$ = P, As, and Sb. 
In the alternating layered structure of $X$ and Pd(dmit)$_2$ molecules, 
 the latter form [Pd(dmit)$_2$]$_2$ dimers arranged 
 in an anisotropic triangular lattice structure 
 within the two-dimensional (2D) planes. 
The cationic $X^+$ molecules are closed-shell and thus do not contribute to 
 the electronic structure near the Fermi level; 
 the anion radical [Pd(dmit)$_2$]$_2$ dimers possess 
 a carrier number of one electron per dimer~\cite{Kato_2004CR,Miyazaki_1999PRB}, 
 governing the electronic properties. 

Their ambient-pressure physical properties can be summarized in 
 a unified phase diagram~\cite{Kanoda_2011ARCMP,Kato_2014BCSJ}, 
 where the horizontal axis is taken as the ratio between two effective interdimer transfer integrals, $t$ and $t'$, 
 based on calculations by the extended H\"{u}ckel method. 
Bonds with $t$ form a square lattice and $t'$ is along one of its diagonals; 
 then, $t'/t=1$ corresponds to the isotropic triangular lattice point. 
The calculated values fall within $t'/t=0.6$ -- 1.0 
 for the members above of $\beta'$-$X$[Pd(dmit)$_2$]$_2$. 
The materials are all insulating, considered as Mott insulators based on dimers (dimer-Mott insulator); 
 then, this ratio is interpreted as a control parameter for the degree of spin frustration. 

In the phase diagram, the transition temperatures of antiferromagnetic (AF) ordering, 
 \tn, are well scaled. 
\tn, which is 40 K for the `end' member $X$ = Me$_4$P with $t'/t \simeq 0.6$, 
 decreases as $t'/t$ increases, 
 and for $X$ = EtMe$_3$Sb ($t'/t \simeq 0.9$), it reaches zero: 
 a spin-liquid state is realized~\cite{Itou_2009PRB,Kanoda_2011ARCMP}. 
Furthermore, 
 $X$ = Et$_2$Me$_2$Sb is situated at a larger $t'/t$ of $\simeq 1$, 
 and exhibits a first-order phase transition at $T=$ 70 K to 
 a charge-ordered (or separated) state~\cite{Nakao_2005JPSJ,Tamura_2005CPL}. 
This state is strongly coupled with lattice distortions, 
 where neutral and divalent dimers are arranged periodically, 
 which we call here charge-lattice order (CLO)  
 to distinguish it from the charge-ordering phenomena in quarter-filled systems~\cite{Seo_2006JPSJ}. 
To summarize, as $t'/t$ increases, 
 the ground state varies as AF $\rightarrow$ spin liquid $\rightarrow$ CLO. 

Recently, 
 a number of first-principles band 
 calculations~\cite{Nakamura_2012PRB,Scriven_2012PRL,Tsumuraya_2013JPSJ,Jacko_2013PRB} 
 based on  
 density functional theory 
 for this family 
 have been performed. 
In Ref.~\citen{Tsumuraya_2013JPSJ}, 
 we estimated the interdimer transfer integrals 
 for the series above, and found that a fully anisotropic triangular lattice 
 is needed to describe the compounds, 
 i.e., the transfer integrals along the three directions are different 
 and vary systematically depending on $X$. 
We also found the existence of charge disproportionation between the two dmit ligands, 
 which is not realized in a monomer Pd(dmit)$_2$ molecule, 
 based on the Kohn-Sham orbitals near the Fermi energy $\ef$ 
 as well as 
 calculations on isolated dimers.  
This phenomenon is 
 associated with the multiorbital nature existing in this system. 
Namely, intradimer hybridization between 
 the highest occupied molecular orbital (HOMO) of one monomer 
 and the lowest unoccupied molecular orbital (LUMO) 
 of the other was suggested, i.e., the existence of HOMO-LUMO mixing. 

The multiorbital nature is known to be a peculiar and important factor 
 in their electronic structures. 
Since the HOMO-LUMO energy gap of the monomer Pd(dmit)$_2$ is relatively small 
 (calculated as $\simeq 0.6$ eV~\cite{Miyazaki_1999PRB}), 
 it was argued that a `HOMO-LUMO inversion' occurs 
 when the strong dimerization observed in solids takes place~\cite{Kato_2004CR,Canadell_1989JPF,Tajima_1991SSC}. 
As a consequence, the bands crossing $\ef$ 
 are considered to consist of HOMOs 
 rather than LUMOs where electrons usually enter 
 in anion radicals. 
This picture is based on the assumption that the HOMO-LUMO mixing is negligible, 
 supported by extended H\"{u}ckel calculations. 
Furthermore, regarding the stability of the CLO state, 
 it was indicated that the HOMO-LUMO inversion is crucial~\cite{Tamura_2004CPL}. 
However, it is unclear to what extent these discussions are applicable 
 when the HOMO-LUMO mixing suggested by our first-principles calculation is taken into account. 

To theoretically analyze such a multiorbital degree of freedom in molecular conductors, 
 some of us and coauthors have recently developed a scheme called the fragment molecular orbital (fMO) 
 approach~\cite{Seo_2008JPSJ,Bonnet_2010JCP,Tsuchiizu_2011JPSJ,Tsuchiizu_2012JCP,Seo_2013JPSJ}. 
It is based on the fact that their frontier molecular orbitals can frequently be described as 
 approximate linear combinations of wave functions localized on several portions of the molecule, i.e., fMOs. 
These fMOs can serve as basis functions of low-energy effective models to describe 
 the electronic properties of multiorbital molecular conductors. 

In this paper, we apply fMO analysis to the family of $\beta'$-$X$[Pd(dmit)$_2$]$_2$. 
Our aim is to address the issues raised above: 
 (i) to theoretically describe the HOMO-LUMO mixing effect in a clear and yet quantitative manner
  from the viewpoint of the fMO scheme 
  by making a correspondence to the HOMO-LUMO inversion picture previously discussed; 
 (ii) to investigate its consequences in the electronic structures, 
 such as the intramolecular charge disproportionation and its relation with the effective dimer model; and 
 (iii) to analyze the ordered phases and their competition observed in experiments, 
 especially determining the characteristic roles of the multiorbital degree of freedom in them. 

In Sect.~\ref{sec2}, we set up our effective multiorbital Hubbard model 
 and discuss the tight-binding parameters 
 obtained by fitting to the results of first-principles band calculations. 
Then, in Sect.~\ref{sec3}, by treating the Coulomb interactions 
 within the mean-field (MF) approximation, we investigate the ground-state properties
 including electron-lattice couplings. 
Sections~\ref{sec4} and \ref{sec5} are devoted to discussions and a summary, respectively. 
In the Appendix, 
 we show electronic structure calculations of the so-called 
$P$2$_1$/$m$ EtMe$_3$P salt, 
 which has a slightly different crystal structure from the $\beta'$-type salts. 


\section{Fragment Model and HOMO-LUMO Mixing}\label{sec2}

\begin{figure}
 \begin{center}
 \vspace*{1em}
  \includegraphics[width=0.3\textwidth]{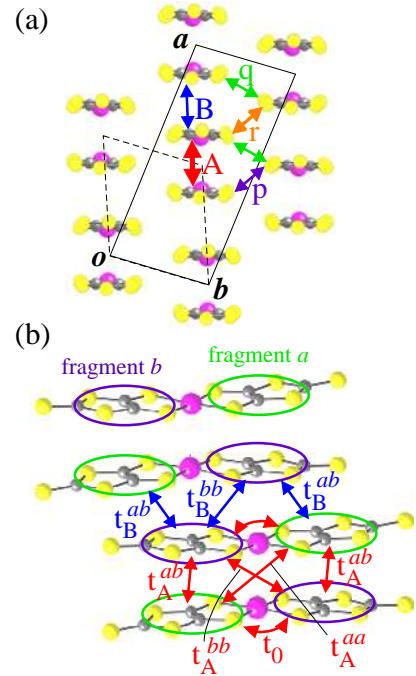}
 \end{center}
 \vspace*{-1em}
 \caption{%
 (Color online) 
 (a) Notation of intermolecular bonds 
 in the 2D layer of Pd(dmit)$_2$ molecules. 
 The conventional unit cell (solid line) and 
 the unit cell for the 2D arrangement (broken line) are shown. 
 (b) Inter-fMO transfer integrals (see text for notation) along the stacking direction. 
 A schematic image of the spatial extent of two types of fMO, $a$ and $b$, is shown. 
 Note that $t_\textrm{B}^{aa}$, not shown in the figure, 
  is included in the fitting described in Sect.~\ref{subsec22} (for example, see Fig.~\ref{fig4}), 
  but is found to be negligibly small. 
 }\label{fig1}
 \vspace*{-1.5em}
\end{figure}

Since the HOMO and LUMO of the isolated Pd(dmit)$_2$ molecule 
 are approximately the bonding and antibonding combinations of the two ligand wave functions, respectively, 
 we consider basis functions of the low-energy effective model 
 as the fMO disproportionated on either of the two dmit ligands. 
This is analogous to the cases previously studied: 
 single-component molecular conductors [$M$(tmdt)$_2$] 
($M$ = Ni, Au, Cu)~\cite{Seo_2008JPSJ,Tsuchiizu_2012JCP,Seo_2013JPSJ}, 
 which consist of metal complex molecules as in the present case, 
 and a purely $p\pi$-orbital system 
 (TTM-TTP)I$_3$~\cite{Bonnet_2010JCP,Tsuchiizu_2012JCP,Tsuchiizu_2011JPSJ}. 
Contributions from the Pd $d$ orbital are known to 
 play important roles~\cite{Kato_2004CR,Miyazaki_1999PRB}, 
 which we will discuss later.  
Then, 
 by noting that the two dmit ligands in each molecule 
 are crystallographically inequivalent, 
 the electronic Hubbard-type Hamiltonian is written as 
\begin{align}
&{\cal H}_0 =\sum_{\langle i,j \rangle, s} t_{ij} \left( c^\dagger_{is} c_{js}^{} + \mathrm{h.c.} \right) \nonumber\\
& \hspace{1cm} + U_{a} \sum_{i \in a} n_{i\uparrow} n_{i\downarrow} 
 + U_{b} \sum_{i \in b} n_{i\uparrow} n_{i\downarrow}.\label{eq1}
\end{align}
Here, 
 $t_{ij}$ denotes the transfer integrals between fMOs, 
 where the sum is taken for inter-fMO pairs $\langle i,j \rangle$ 
 including both intramolecular and intermolecular bonds, 
 and $c_{is}$ ($c^\dagger_{is}$) denotes the electron annihilation (creation) operator 
 for the fMO site index $i$ and spin $s= \uparrow$ or $\downarrow$. 
The next two terms are the on-site Coulomb repulsion 
 for the two types of fMO site corresponding to the two ligands on each molecule
 denoted as $a$ and $b$ (see Fig.~\ref{fig1}), $U_{a}$ and $U_{b}$, respectively. 
The density operators are defined in the normal-ordered form, 
 i.e., $n_{is} = c^\dagger_{is} c_{is} - \bar{n}_\mu$ for $i\in\mu=a$ or $b$, where
 $\bar{n}_\mu$ is the mean electron density per spin on the $\mu$-fMO site~\cite{Tsuchiizu_2012JCP}. 
The average electron density of each molecule is 5/2 including the HOMO and LUMO, 
 corresponding to the two fragments; 
 then, $2(\bar{n}_a + \bar{n}_b) = 5/2$; 
 the mean electron density per fragment is 5/4.

The model here is simplified by neglecting possible additional terms. 
First, since $a$- and $b$-fMOs are inequivalent, 
 there should be a difference in their one-body `orbital energies'. 
We expect it to be small owing to the nearly symmetric structure of the Pd(dmit)$_2$ molecule, 
 and thus do not include it. 
It should mainly affect the difference in the electron occupancy between $a$ and $b$, 
 which is also affected by the transfer integrals, as we will see below. 
In this sense, the effect of different orbital energies is 
 partly included in the parameters $t_{ij}$. 

Two-body interaction terms other than $U_{a}$ and $U_{b}$, 
  such as inter-fMO Coulomb repulsions and exchange interactions, 
  are also not included here. 
In the fMO scheme, 
 even if the wave functions are extended on certain portions of the molecule,
 an intuitive description based on the tight-binding picture works well. 
Namely, the inter-fMO Coulomb repulsions scale well with 
 the inverse of the distance between `center positions' of fMOs, 
 and the exchange couplings are smaller than 
 the direct terms~\cite{Tsuchiizu_2012JCP,Tsuchiizu_2011JPSJ}. 
We adopt such calculations for other molecular materials, 
 and first consider $U_{a}$ and $U_{b}$, naturally expected to be the largest. 
In the calculations described in Sect.~\ref{subsec32}, 
 we also performed calculations including 
 the inter-fMO Coulomb repulsions scaled by their inverse distances.

In Fig.~\ref{fig1}, 
 the 2D structure of a Pd(dmit)$_2$ layer 
 and a schematic view of the two types of fMO are shown. 
The notation for the intermolecular bonds [Fig.~\ref{fig1}(a)] 
 is taken from the literature~\cite{Kato_2004CR}. 
For example, bond `A' represents the intradimer bond.
As shown in Fig.~\ref{fig1}(b), 
 the intermolecular transfer integrals are denoted as 
 $t_\alpha^{\mu\nu}$, where $\alpha$ is the bond index 
 and $\mu$ and $\nu$ take $a$ or $b$ indicating the fMO pair; 
 the intramolecular transfer integral is written as $t_0$. 
The interlayer parameters are included in the fitting procedure 
 but found to be small for the $\beta'$-type salts. 
(In fact, this is not the case in the $P$2$_1$/$m$ EtMe$_3$P salt; see the Appendix.) 

\subsection{Isolated dimer; HOMO-LUMO inversion}\label{subsec21}

\begin{figure}
 \vspace*{1em}
 \begin{center}
  \includegraphics[width=0.47\textwidth]{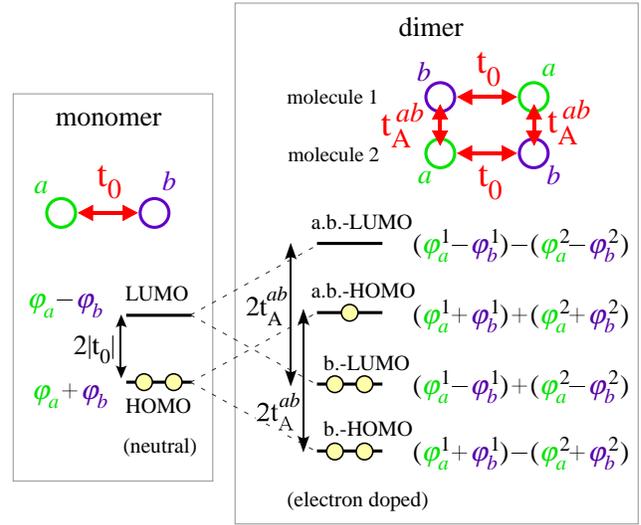}
 \end{center}
 \caption{%
 (Color online) HOMO-LUMO inversion picture based on the fMO scheme. 
We show energy levels of an isolated monomer and dimer with no HOMO-LUMO mixing. 
The wave functions are also indicated, where renormalization factors are omitted. 
The electron configurations are for a neutral monomer 
 and for an electron-doped dimer realized in the crystals. 
`b.' and `a.b.' stand for the bonding and antibonding combinations of the molecular orbitals, respectively. 
 }\label{fig2} 
 \vspace*{-1em}
\end{figure}

First, let us consider an isolated dimer with two sets of $a$- and $b$- fMO sites (Fig.~\ref{fig2}), 
 and clarify the relationship between the previous discussions based on molecular orbitals 
 and our fMO scheme. 
In the latter, there are four types of intradimer transfer integrals, 
 $t_0$, $t_\textrm{A}^{ab}$, $t_\textrm{A}^{aa}$, and $t_\textrm{A}^{bb}$, 
 as shown in Fig.~\ref{fig1}(b). 
Let us fix signs as $t_0 < 0$ and $t_\textrm{A}^{ab} > 0$
 in accordance with 
the discussions hereafter. 
The energy levels of a monomer with two fMO sites, $a$ and $b$, 
 connected by $t_0$ are $\pm t_0$. 
The bonding and antibonding combinations 
 of ligand wave functions correspond to 
 the HOMO [$t_0$ ($<0$)] and LUMO [$-t_0$ ($>0$)] of a neutral molecule; 
 the HOMO-LUMO gap is $2|t_0|$. 
Then, if we consider only $t_\textrm{A}^{ab}$ as the intermolecular transfer integrals 
 as in Fig.~\ref{fig2}, 
 the two levels respectively split and result in four energy levels, 
 of $\pm t_0 \pm t_\textrm{A}^{ab}$. 
The wave functions are 
 the interdimer bonding combination of the HOMO pair ($t_0-t_\textrm{A}^{ab}$), 
 antibonding HOMO pair ($t_0+t_\textrm{A}^{ab}$), 
 bonding LUMO pair ($-t_0-t_\textrm{A}^{ab}$), 
 and antibonding LUMO pair ($-t_0+t_\textrm{A}^{ab}$). 
The HOMO-LUMO inversion occurs when $|t_0| < t_\textrm{A}^{ab}$, 
 and the antibonding HOMO pair becomes singly occupied in the anion radical salt. 

In this situation, the overlap between 
 the HOMO of one molecule and the LUMO of the other becomes zero; 
 HOMO-LUMO mixing does not occur. 
An important point is that, 
 if $a$ and $b$ are equivalent, 
 the relation $t_\textrm{A}^{aa} = t_\textrm{A}^{bb}$ holds, 
 and then even when these are finite, 
 the zero hybridization condition still holds. 
In such a case, 
  all the wave functions have equal weight of the four fMOs, as indicated in Fig.~\ref{fig2}. 
The anti-bonding HOMO pair level, where the doped electron enters, 
 has a wave function proportional to 
 $\varphi_1^a + \varphi_1^b + \varphi_2^a + \varphi_2^b$, 
 where $\varphi_i^\mu$ denotes the fMO wave function on molecule $i =$ 1 or 2 and fMO $\mu = a$ or $b$. 
These properties should be maintained even under interaction, 
 as long as an isolated system is considered where no symmetry breaking occurs. 

We will see in the following that this situation is actually violated  
 in the crystal. 
The mirror symmetry for each molecule is broken,  
 and we find that $t_\textrm{A}^{aa}$ and $t_\textrm{A}^{bb}$ are markedly different. 
This is consistent with the crystal structure where  
 relatively large bending of the molecule is observed experimentally~\cite{Kato_2013Crystals}.

\subsection{Tight-binding parameters}\label{subsec22}

\begin{figure}
 \begin{center}
  \includegraphics[width=0.5\textwidth,clip]{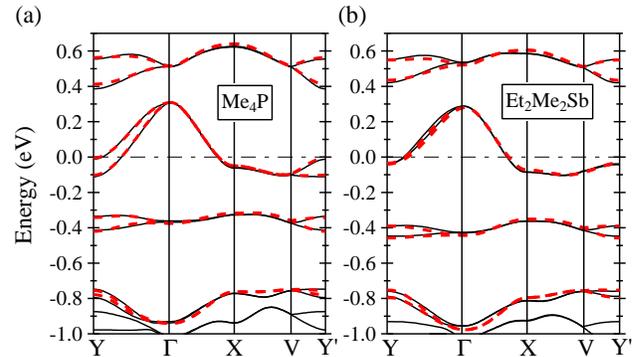}
 \end{center}
 \caption{%
 (Color online) 
 Electronic band structures near $\ef$ for $\beta'$-$X$[Pd(dmit)$_2$]$_2$ 
 with (a) $X=$ Me$_4$P and (b) Et$_2$Me$_2$Sb. 
 Solid and broken lines are the first-principles~\cite{Tsumuraya_2013JPSJ} 
 and fitted tight-binding bands, respectively. 
 }\label{fig3}
 \vspace*{-1em}
\end{figure}

The tight-binding parameters $t_\alpha^{\mu\nu}$ 
 are obtained by numerically fitting the band structures 
 by the first-principles calculations described in Ref.~\citen{Tsumuraya_2013JPSJ}. 
In Fig.~\ref{fig3}, 
 we show the first-principles bands near $\ef$ 
 together with the fitted tight-binding dispersions 
 for the two end members in the phase diagram explained in Sect.~\ref{sec1}, 
 i.e., $X=$ Me$_4$P and Et$_2$Me$_2$Sb, as examples. 
The primitive cell consists of four Pd(dmit)$_2$ molecules, that is, eight fMO sites; 
 owing to the solid crossing column structure, 
 two crystallographically equivalent layers with different stacking directions exist, 
 each containing two molecules in the primitive cell. 
The eight bands that we fit originate from the eight fMOs, 
 or equivalently, four sets of HOMO and LUMO. 
Although the bands at the bottom are close to other bands, 
 one can see that the fitting near $\ef$ is satisfactory. 

\begin{figure}
 \begin{center}
  \includegraphics[width=0.5\textwidth]{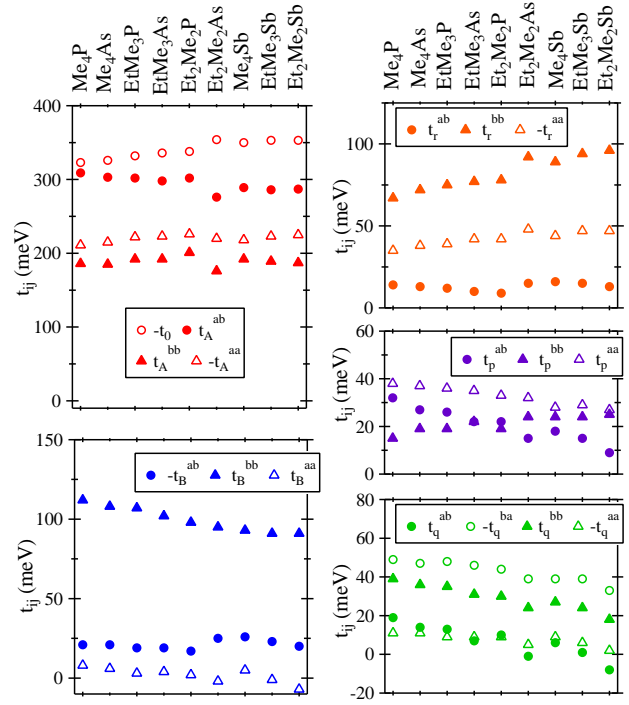}
 \end{center}
 \vspace*{-2em}
 \caption{%
 (Color online) 
Transfer integrals $t_{ij}$ obtained by fitting the first-principles band structures 
 for the series of compounds $\beta'$-$X$[Pd(dmit)$_2$]$_2$. 
The list of $X$ is shown at the top. 
For the notations of specific $t_{ij}$, represented as $t_\alpha^{\mu\nu}$ 
 ($\alpha$: bond index and $\mu, \nu$: $a$ or $b$), see Fig.~\ref{fig1}(b). 
Values of the interlayer $t_{ij}$ are not shown, which are on the order of 1 meV.
 }\label{fig4} 
 \vspace*{-1em}
\end{figure}

In Fig.~\ref{fig4}, the obtained parameters are plotted, 
 where the order of $X$ on the horizontal axis is adopted from 
 that in the experimental phase diagram~\cite{Kato_2014BCSJ}. 
Interlayer parameters are small, as noted above ($\sim$ 1 meV), 
 and therefore not shown in the figure. 
The overall trend shows a systematic variation, 
 which is common to the extended H\"{u}ckel calculations. 
However, 
 a newly found feature is seen in the intradimer transfer integrals 
 shown in the upper left panel of Fig.~\ref{fig4}. 
All four types of $t_{ij}$ discussed in Sect.~\ref{subsec21} are of the same order.
One can see that $|t_0| \gsim t_\textrm{A}^{ab}$, 
 different from the condition mentioned above in the HOMO-LUMO inversion picture, 
 and more importantly,  
 $t_\textrm{A}^{aa} \neq t_\textrm{A}^{bb}$; they even have opposite signs. 
This gives rise to appreciable hybridization in the original molecular orbital picture, 
 and also to a discrepancy from the equal-weight situation in Sect.~\ref{subsec21}, 
 as we discuss in the following. 
We note again that these values are obtained by assuming that the orbital energies 
 are equal, and therefore should be considered as effective values. 

When we calculate the eigenstates of an isolated dimer 
with two pairs of $a$- and $b$-fMO sites 
 (a four-site problem) 
 with the parameters obtained by the fitting, 
 large disproportionated weights between $a$- and $b$-fMOs are seen 
 for the second highest energy level, 
 where the singly occupied electron is expected to enter (see Fig.~\ref{fig2}). 
For example, 
 the ratio of the coefficient of $\varphi^b$ to that of $\varphi^a$ 
 in the obtained wave function is 28.4 (6.4) 
 for the parameters for $X=$ Me$_4$P (Et$_2$Me$_2$Sb). 
Namely, they are almost completely disproportionated. 
Then, the electron occupation that requires summation including the levels below $\ef$,  
 becomes almost 1.5 and 1 for the $a$ and $b$ fMO sites, respectively  
(the level below has a large weight of $\varphi^a$). 
This is for an isolated dimer. 

The full tight-binding model including all $t_{ij}$ 
 gives electron occupations of 5/4 $\pm \ \delta$ with $\delta= 0.08$ -- 0.09
 for $a$ ($+ \delta$) and $b$ ($- \delta$) fMOs, respectively. 
This reduction of disproportionation compared with the four-site problem above ($\delta=0.25$)
 is due to the band effect from the interdimer transfer integrals. 
Note that, 
 in our approach here, 
 we cannot qualitatively discuss the charge disproportionation 
 at a microscopic atomic scale  
 since the fMOs are not derived but introduced a priori; 
 the disproportionation may be related to the imbalance of spatial extensions of the fMO. 
Nevertheless, 
 such a charge disproportionation was also found in our previous ab initio
 study~\cite{Tsumuraya_2013JPSJ}, 
 supporting the existence of charge imbalance. 

Although it is not straightforward to elucidate the microscopic origin of 
 such large hybridization from our fitting procedure, 
 let us discuss some intuitive considerations 
 on the difference of the signs in $t_\textrm{A}^{aa}$ and $t_\textrm{A}^{bb}$. 
If we assume wave functions having $p$-wave symmetry with nodes on the molecular plane for all fMOs, 
 negative $t_0$ and positive $t_\textrm{A}^{ab}$ are naively expected to lead to 
 positive values for both $t_\textrm{A}^{aa}$ and $t_\textrm{A}^{bb}$. 
The discrepancy from such a naive picture 
 might originate from the contribution from the Pd $d$ orbital,  
 whose importance has been discussed from different aspects~\cite{Kato_2004CR,Miyazaki_1999PRB}. 
The $d$ orbital at the Pd site mixes with 
 the ligand $p\pi$ orbital 
 and can give a different sign on the opposite ligand side. 
Therefore, a negative value is 
 expected if the weight of the $d$ orbital is large. 
Or, equivalently, higher-order hopping processes including $p\pi$ and $d$ orbitals can 
 alter the sign of effective transfer integrals in general. 

The interdimer bonds, on the other hand, 
 show transfer integrals smaller than those of the intradimer bonds. 
Among them, the largest is $t_\textrm{B}^{bb} \sim $ 100 meV, in the stacking direction. 
One can see in Fig.~\ref{fig1}(b) that the $b$-fMOs are close in distance 
 along the B bond, 
 which is the reason for the large value. 
A characteristic feature is that 
 most of the them taking large values decrease from left to right by varying $X$, 
 but $t_\textrm{r}^{bb}$, which is also as large as 70 -- 100 meV, increases.
This is common to the extended H\"{u}ckel calculations; 
 it has been pointed out that a distortion within the Pd(dmit)$_2$ molecule 
 accounts for such behavior~\cite{Kato_2013Crystals}. 

\subsection{Effective dimer model}\label{subsec23}

Let us discuss the consequence of the results shown above 
 on the effective half-filled model frequently discussed to describe the 
 electronic properties of this system, especially the spin-liquid state 
 in $X=$ EtMe$_3$Sb. 
Since the band crossing $\ef$ is half-filled, 
 one can consider a tight-binding model based on a basis set 
 by which the band is constructed. 
This corresponds to taking dimers as units, i.e., the lattice points for the model. 
In single-orbital molecular conductors with a dimerized structure, 
 the basis function is approximated by the bonding or antibonding 
 combination of the molecular orbital 
 where the carriers exist, 
 from which the parameters of the model can be derived~\cite{Kino_1996JPSJ}. 
For example, in $\kappa$-(BEDT-TTF)$_2X$, it is the antibonding HOMO pair 
 where the holes are introduced
owing to the cationic radical [(BEDT-TTF)$_2$]$^+$ dimers. 
In $\beta'$-$X$[Pd(dmit)$_2$]$_2$, 
 based on the extended H\"{u}ckel calculations, 
 the parameters have been derived by taking the antibonding HOMO pair~\cite{Kato_2004CR} 
 as a result of the HOMO-LUMO inversion. 
As mentioned in Sect.~\ref{subsec21}, 
 this wave function has an equal weight for all fMOs within the dimer, 
 which is not the case in the results obtained in Sect.~\ref{subsec22}. 

\begin{figure}
 \vspace*{1em}
 \begin{center}
  \includegraphics[width=0.4\textwidth]{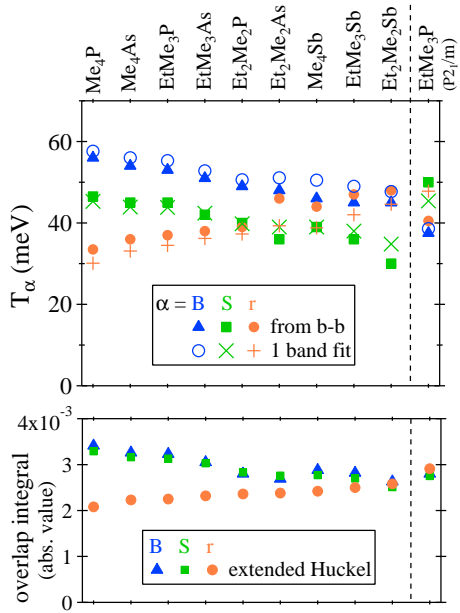}
 \end{center}
 \vspace*{-2em}
 \caption{%
 (Color online) 
Effective interdimer transfer integrals 
 derived from the transfer integrals between the $b$-fMOs shown in Fig.~\ref{fig4} 
  (`from b-b') and 
 those obtained by fitting the half-filled band crossing $\ef$ (`1-band fit')~\cite{Tsumuraya_2013JPSJ}. 
Overlap integrals calculated by the extended H\"{u}ckel approximation are plotted in the lower panel; 
 in this scheme, the transfer integrals are approximated as proportional to the overlaps. 
Results for the $P$2$_1$/$m$ EtMe$_3$P salt are also plotted for reference (see Appendix). 
 }\label{fig5}
 \vspace*{-1em}
\end{figure}

Here, instead, as an attempt, 
 we take the basis function of the dimer model as the wave function 
 completely disproportionated on the $b$-fMO, i.e., 
 $\frac{1}{\sqrt{2}}(\varphi_1^b+\varphi_2^b)$, 
 and estimate the effective interdimer transfer integrals from the fitted values in Fig.~\ref{fig3}. 
This is based on the results mentioned in the previous subsection for the isolated dimer. 
Then, the three types of interdimer transfer integral forming an anisotropic triangular lattice, 
 $T_\textrm{B}$, $T_\textrm{r}$, and $T_\textrm{S}$, 
 can be evaluated from $t_\alpha^{bb}$, as 
 $T_\textrm{B}=t_\textrm{B}^{bb}/2$, $T_\textrm{r}=t_\textrm{r}^{bb}/2$, 
 and $T_\textrm{S}=t_\textrm{q}^{bb}+t_\textrm{p}^{bb}/2$. 
The values are plotted in Fig.~\ref{fig5}, 
 together with the corresponding transfer integrals evaluated in Ref.~\citen{Tsumuraya_2013JPSJ}, 
 obtained by directly performing the fitting to the half-filled band. 
One can see that the values evaluated by different methods are in agreement, 
 at least at a semiquantitative level. 
As pointed out in Ref.~\citen{Tsumuraya_2013JPSJ}, 
 the three parameters are rather different from each other, 
 in contrast to the extended H\"{u}ckel parameters 
 plotted in the lower panel of Fig.~\ref{fig5} giving $T_\textrm{B} \simeq T_\textrm{S}$.
From the results here, 
 we can deduce that the disproportionation between the ligands, 
 which is a consequence of the HOMO-LUMO mixing, 
 accounts for the revision of the parameters of the effective dimer model. 

\section{Broken-Symmetry States}\label{sec3}

Now, we take into account the effects of interactions and investigate broken-symmetry states 
 based on the fMO model introduced in Sect.~\ref{sec2} [Eq.~(\ref{eq1})].  
Here, we consider 2D models 
 where the interlayer couplings are neglected. 
The on-site Coulomb interactions are treated within the MF approximation, 
 $n_{i\uparrow}n_{i\downarrow} \rightarrow \langle n_{i\uparrow} \rangle n_{i\downarrow} 
 + n_{i\uparrow} \langle n_{i\downarrow} \rangle
 - \langle n_{i\uparrow} \rangle \langle n_{i\downarrow} \rangle$, 
 and we set $U_a = U_b \equiv U$ for simplicity. 
We further take into account  
 intersite Coulomb interactions and 
 couplings to lattice degrees of freedom in Sect.~\ref{subsec32}. 
Self-consistent solutions are obtained for a 2 $\times$ 2 supercell 
 of the conventional unit cell of 4 molecules = 8 fMO sites. 

\subsection{Antiferromagnetic states}\label{subsec31}

\begin{figure}
 \vspace*{1em}
 \begin{center}
  \includegraphics[width=0.4\textwidth]{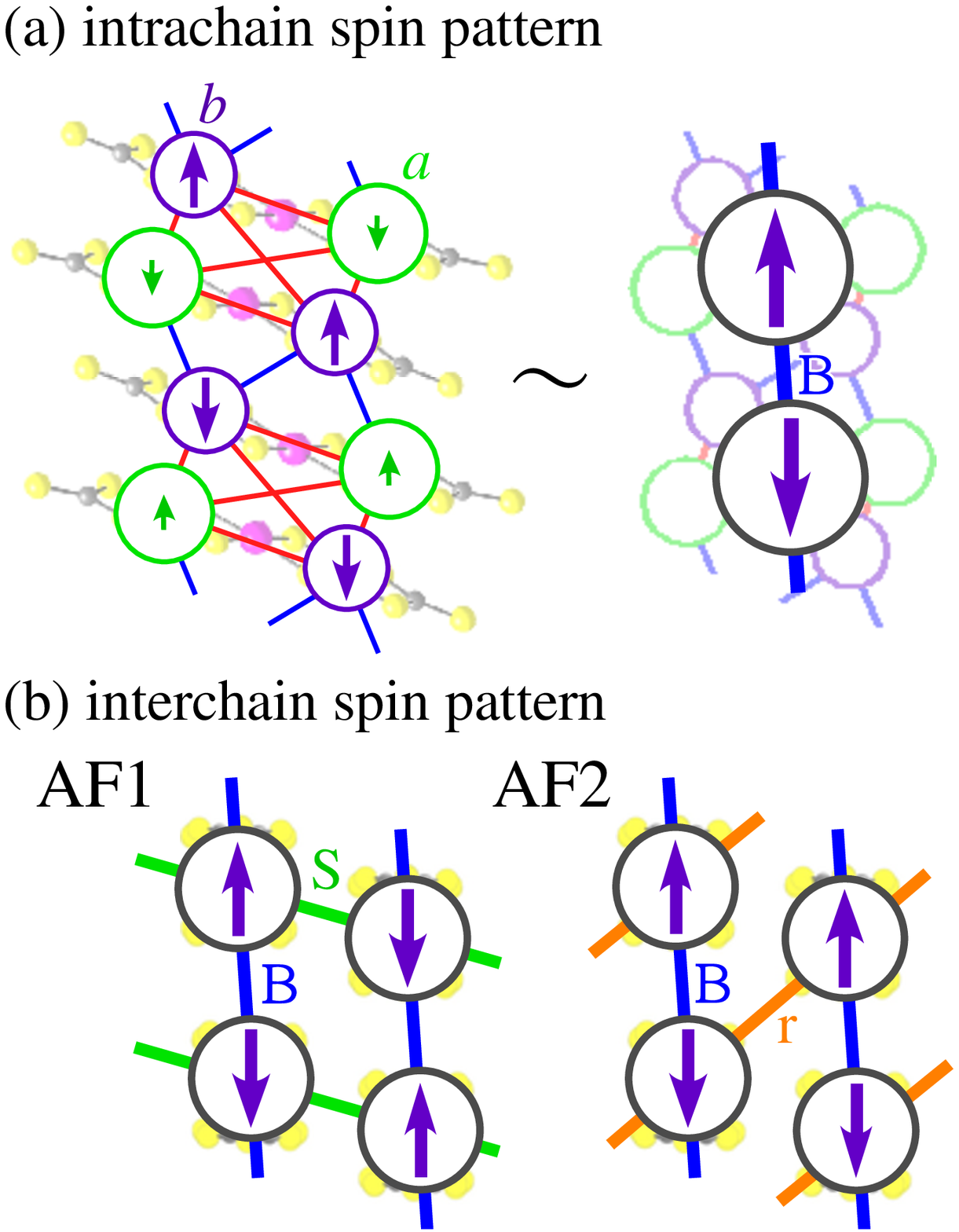}
 \end{center}
 \vspace*{-1em}
 \caption{%
 (Color online) 
Schematic views of stable mean-field solutions for the effective Hubbard model. 
AF patterns are shown by arrows 
 representing up/down spins 
 while their sizes and the circles 
 represent spin and charge densities, respectively. 
 (a) Stacking (intrachain) direction: fMO sites with inter-fMO bonds are indicated on the left, 
  while the dimer AF picture is shown on the right. 
 (b) AF alignments in the 2D plane are shown for patterns AF1 (left) and AF2 (right). 
 The dimer picture is adopted where the interdimer bonds with the staggered spin direction are indicated. 
}\label{fig6}
 \vspace*{-1em}
\end{figure}

When $U$ is increased, for all the parameters obtained for different $X$ in Sect. \ref{subsec22}, 
 AF insulating states become stabilized 
 over the paramagnetic metallic state that is stable when $U$ is small. 
The lowest-energy solutions 
 show two different AF patterns at large $U$ ($\simeq$ 1.0 eV) 
 depending on the parameters for different $X$. 
The pattern in the stacking direction along the A and B bonds is common, 
 as shown in Fig.~\ref{fig6}(a). 
There are two types of dimer, 
 where the net magnetic moments show a staggered alignment as schematically shown in the figure; 
 then, at a first approximation the so-called dimer AF state is stabilized. 
The spin moments are mostly carried by the $b$-fMO, 
 but the $a$-fMO within the same dimer shows opposite 
 spin directions: 
 an intramolecular AF state. 
This is not anticipated from the single-MO picture; 
 therefore, it is a consequence of the HOMO-LUMO mixing. 
We note that an intramolecular AF pattern was also stabilized in 
 calculations for another multiorbital system, i.e., 
 the single-component molecular conductor Au(tmdt)$_2$~\cite{Ishibashi_2005JPSJ,Seo_2008JPSJ}. 

The two AF patterns differ in the interstack direction, 
 as shown in Fig.~\ref{fig6}(b). 
AF1 shows a staggered alignment along the interdimer S bonds, 
 while in AF2, the direction of moments is staggered along the r bonds. 
In the dimer picture where each dimer carries effective spin $S=$ 1/2, 
 these patterns correspond to N\'{e}el states 
 with the strongest exchange coupling along the B bonds, 
 followed by that in the S (r) bonds for AF1 (AF2). 

\begin{figure}
 \vspace*{1em}
 \begin{center}
  \includegraphics[width=0.5\textwidth]{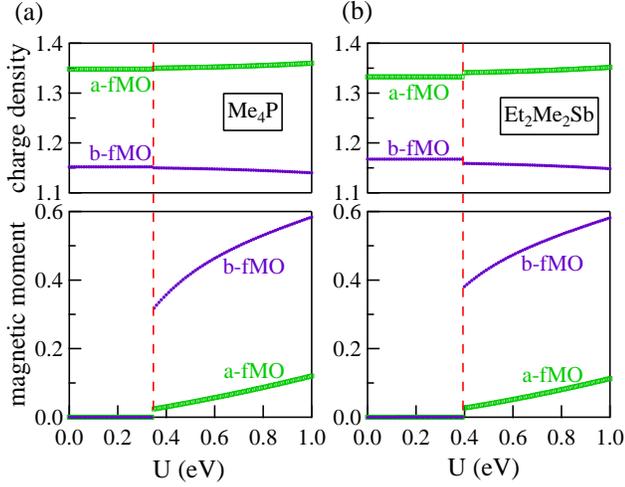}
 \end{center}
 \vspace*{-1em}
 \caption{%
 (Color online) 
$U$-dependences of charge densities (upper panels) and 
 absolute values of magnetic moment (lower panels) 
 on $a$- and $b$-fMO sites 
 for parameters for $\beta'$-$X$[Pd(dmit)$_2$]$_2$ 
 with (a) $X=$ Me$_4$P and (b) Et$_2$Me$_2$Sb. 
Note that 
 the magnetic moments on $a$- and $b$-fMOs within each dimer show opposite signs, 
 and the AF patterns are different for the two cases (see Fig.~\ref{fig6} and text). 
}\label{fig7}
 \vspace*{-1em}
\end{figure}

In Fig.~\ref{fig7}, the $U$-dependences of charge and spin density on each site are plotted 
 for the end members $X=$ Me$_4$P (pattern AF1) 
 and Et$_2$Me$_2$Sb (pattern AF2). 
As $U$ is increased, a first-order phase transition occurs 
 from the paramagnetic metallic state to the AF insulating state 
 at $U_\textrm{c}=0.35$~eV (0.39 eV) 
 in the case of the parameters for $X=$ Me$_4$P (Et$_2$Me$_2$Sb). 
The difference in $U_\textrm{c}$ does not originate from 
 the difference in their bandwidth which actually shows a decrease when varying the cation 
 from left to right in the phase diagram~\cite{Tsumuraya_2013JPSJ}. 
A possible origin is that the materials on the left are more anisotropic 
 in terms of their interdimer transfer integrals, 
 which leads to a larger degree in the nesting of the Fermi surface, 
 relatively stabilizing the AF state. 
The result that the $b$-fMO carries most of the spin moment 
 is consistent with the discussions in Sect. \ref{sec2}, that is,  
 the level where the singly occupied electron enters
 is mainly characterized by the $b$-fMO. 
Another point to be noticed is that the magnetic moment on the $b$-fMO 
 exceeds 0.5 at sufficiently large $U$. 
This is also peculiar compared with the single-MO cases. 
 For example,
 in a similar MF calculation for $\kappa$-(BEDT-TTF)$_2X$~\cite{Kino_1996JPSJ}, 
 the magnetic moment on each BEDT-TTF site is always less than 0.5 (1 $\mu_\textrm{B}$ per dimer). 

At $U\simeq$ 1.0 eV, 
 AF1 is favored for the parameters of $X=$ Me$_4$P -- Et$_2$Me$_2$P, 
 whereas AF2 is favored for $X=$ Et$_2$Me$_2$As -- Et$_2$Me$_2$Sb. 
This is consistent with the tight-binding parameters in the dimer model, 
 as discussed in Sect.~\ref{subsec23}, where the crossover from $T_\textrm{S} > T_\textrm{r}$ to 
 $T_\textrm{S} < T_\textrm{r}$ occurs around $X=$ Et$_2$Me$_2$As;   
 the naive estimation of spin exchange couplings 
 proportional to the square of interdimer transfer integrals 
 is consistent with the stable patterns. 
Such competing AF states are  
 a consequence of spin frustration appearing in the MF level. 
In fact, the MF calculations show keen competition 
 between different patterns in the intermediate-$U$ region 
 for members located near $T_\textrm{S} \simeq T_\textrm{r}$, 
 but we do not present detailed results here  
 since in such a region, quantum fluctuations, neglected in the MF approximation, 
 will be crucial. 

\subsection{Charge-lattice-ordered state}\label{subsec32}

Next, we consider electron-lattice couplings and show that they stabilize CLO states. 
Experimentally, 
 large lattice distortions are observed~\cite{Nakao_2005JPSJ}, 
 in terms of both intermolecular displacements and intramolecular deformations.  
Therefore, we take into account such an effect by considering 
 Peierls- and Holstein-type couplings, 
 given, in addition to Eq.~(\ref{eq1}), as 
\begin{align}
&{\cal H}_{\rm P} =\!\sum_{\langle i j\rangle, s} t_{ij} \alpha_{ij} u_{ij} \left(c_{is}^{\dag}c_{js}+{\rm h.c.}\right)
 +\frac{K_{\rm P}}{2}\sum_{\langle ij\rangle}u_{ij}^2,\label{eq2}\\
&{\cal H}_{\rm H} =-\sum_{i}v_i n_i+\frac{K_{\rm H}}{2}\sum_{i}v_i^2,\label{eq3}
\end{align}
 where $u_{ij}$ and $v_{i}$ are renormalized lattice distortions 
 treated here as classical values; 
their spring constants 
 are $K_{\rm P}$ and $K_{\rm H}$, respectively. 
The Peierls distortions are treated only for the intradimer bonds~\cite{Tamura_2004CPL}, 
 i.e., $u^{\mu\nu}_A$ and $u_0$, 
 corresponding to the bonds with $t^{\mu\nu}_A$ and $t_0$, respectively; 
 otherwise, $u_{ij}=0$. 
 $\alpha_{ij}$ is a coupling constant set as 1 and 0.5 for the A and intramolecular bonds, respectively. 
These are adopted from experimental results~\cite{Nakao_2005JPSJ} 
 that the intradimer bonds show large displacements 
 and from the fact that the intramolecular bonds are harder in general 
 owing to the structural stability of covalent bonds.  
The Holstein coupling reflects 
 the changes in the interatomic bond length
 within the Pd(dmit)$_2$ molecule~\cite{Nishioka_2013JPSJ}. 
Using the Hellman-Feynman theorem under 
 the constraint $\sum_{\langle ij \rangle} u_{ij}=0$, 
 we can obtain $u_{ij}$ and $v_i$ $(=\langle n_i \rangle/K_{\rm H})$ self-consistently
 using the expectation values for bond operators and charge densities~\cite{Yoshimi_2012PRL}. 
In the following, we show results for the fixed condition 
 $K_{\rm H} = 4K_{\rm P} \equiv K$  
 and vary $K$; 
 the strength of the coupling 
(the `softness' of the lattice) 
is controlled by $1/K$. 
 
\begin{figure}
 \vspace*{1em}
 \begin{center}
  \includegraphics[width=0.42\textwidth]{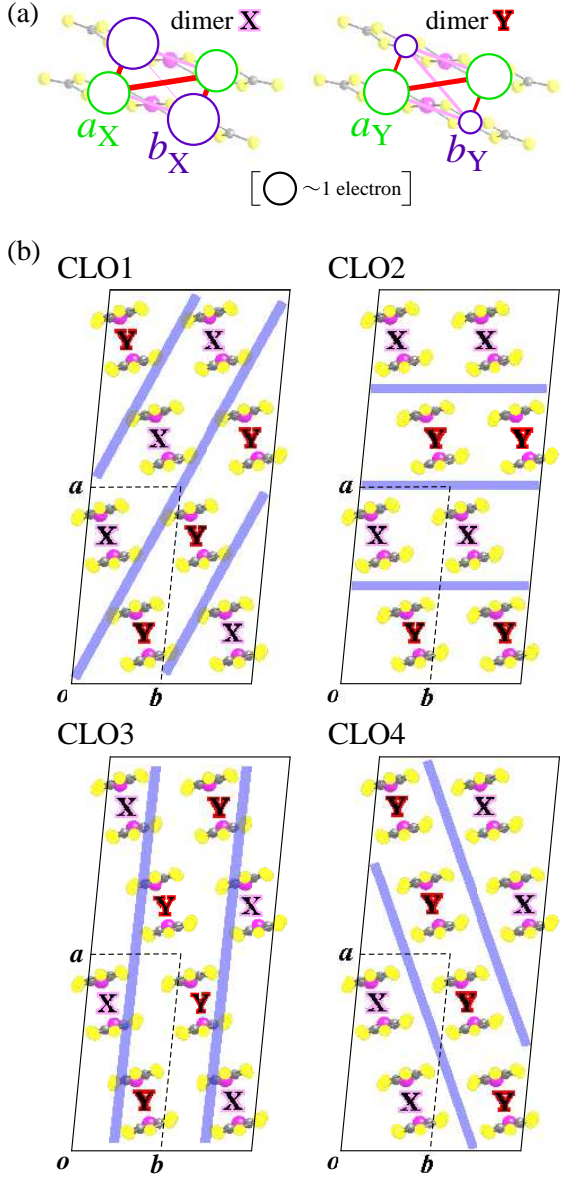}
 \end{center}
 \vspace*{-1em}
 \caption{%
 (Color online) 
Schematic views of charge-lattice-ordered (CLO) states. 
 (a) Two types of [Pd(dmit)$_2$]$_2$ dimer, X and Y. 
 The sizes of the circles and the thicknesses of the bonds represent 
 the charge densities and Peierls lattice distortions, respectively. 
 The dark (light) bonds indicate where typically $u_{ij} > 0$ ($<0$); 
 the size of the circle for one electron is shown as a guide. 
(b) CLO alignments in the 2D plane. Four patterns are obtained. 
 The supercell size that we used and the conventional unit cell are shown as 
 solid and dotted lines, respectively.
}\label{fig8}
 \vspace*{-1em}
\end{figure}

Furthermore, we also consider the intersite Coulomb repulsions, 
 which affect the stability of the CLO state,  
 represented as
\begin{align}
&{\cal H}_V = \sum_{\langle ij \rangle} V_{ij} n_i n_j \ ,\label{eq4}
\end{align}
 where $n_i = n_{i\uparrow} + n_{i\downarrow}$, 
 to which we also apply the MF approximation
 as 
 $n_{i}n_{j} \rightarrow \langle n_{i} \rangle n_{j} 
 + n_{i} \langle n_{j} \rangle
 - \langle n_{i} \rangle \langle n_{j} \rangle$. 
The Coulomb energies are evaluated as 
 proportional to the inverse of the inter-fMO distance, 
 i.e., $V_{ij} \propto 1/r_{ij}$, 
 following the results for the case of (TTM-TTP)I$_3$ 
 discussed in Ref.~\citen{Tsuchiizu_2011JPSJ}. 
Here, 
 we define $1/r_{ij}$ by the average of the inverse distances between two sulfur and two carbon atoms 
 surrounding the Pd atom in each fMO, 
 since the HOMO and LUMO are largely weighted on them. 
We have performed calculations for the cases of $X$ = Me$_4$P and Et$_2$Me$_2$Sb, 
 whose values will be used below, 
 for all the bonds within the 2D layer that we took into account in Sect.~\ref{sec2}. 
For both $X$, 
 $V_{ij}$ for the $a$-$b$ fMO pair along the intradimer A bond becomes the largest 
 (the bond with $t_\textrm{A}^{ab}$); 
 we set it as $V$ as a parameter 
 and fix the ratio with that to the other bonds.  
The next largest is for the $a$-$b$ pair along B of $\simeq 0.9V$, 
 and the next is $\simeq 0.8V$ for the intramolecular bond;  
 the others range from approximately $0.7V$ to $0.4V$. 

We find four patterns of CLO states as self-consistent solutions. 
In all of them, two types of [Pd(dmit)$_2$]$_2$ dimer  (= fMO tetramer)
 denoted as dimer X and dimer Y following Ref.~\citen{Nakao_2005JPSJ}, 
 as shown in Fig.~\ref{fig8}(a), exist in a 1:1 ratio; 
 they are arranged periodically. 
In Fig.~\ref{fig8}(b), 
 the obtained patterns CLO1 -- CLO4 are shown; 
 they correspond to the patterns 
 in Ref.~\citen{Tamura_2005SM}, 
 which were considered to estimate their Madelung energies.~\cite{noteTamura}   
Pattern CLO3 is the one realized in $\beta'$-Et$_2$Me$_2$Sb[Pd(dmit)$_2$]$_2$, 
 revealed by X-ray crystal structure analysis~\cite{Nakao_2005JPSJ}. 

\begin{figure}
 \vspace*{2em}
 \begin{center}
  \includegraphics[width=0.5\textwidth]{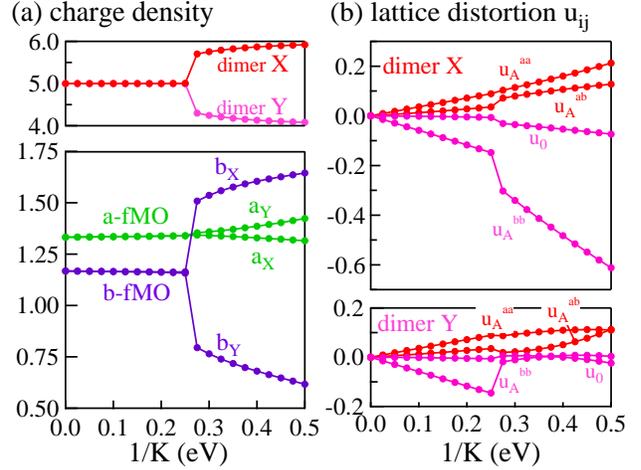}
 \end{center}
 \vspace*{-1em}
 \caption{%
 (Color online) 
$1/K$-dependences of charge densities (a) 
 and Peierls lattice distortions $u_{ij}$ (b) 
 for the transition between the paramagnetic metal and the CLO insulating state. 
The parameters are chosen for $X=$ Et$_2$Me$_2$Sb, 
 with fixed $U=0.3$ eV and $V=0$. 
The ratio between the Holstein and Peierls couplings is fixed as $K_{\rm H} = 4K_{\rm P} \equiv K$; 
 for details, see text. 
}\label{fig9}
 \vspace*{-1em}
\end{figure}

In Fig.~\ref{fig9}, 
 the $1/K$-dependences of charge densities, 
 $\langle n_i \rangle$ ($=K_{\rm H} v_i$), 
 and the Peierls lattice distortions [$u_{\rm A}^{\mu\nu} (\mu,\nu=a,b)$, $u_0$] 
 are plotted  
 to present the character of the CLO state. 
As a typical example, the case for 
 $X$ = Et$_2$Me$_2$Sb
 for fixed $U=$ 0.3~eV and $V=0$ is shown. 
At $1/K=0$, 
 the lattice degree of freedom is in the rigid limit 
 and the system is a paramagnetic metallic state, as described in Sect.~\ref{subsec31}. 
As $1/K$ is increased
 to above $(1/K)_{\rm c} = 0.27$ eV, CLO2 becomes the lowest-energy state. 
There, the system is insulating and shows no magnetization; 
 a first-order phase transition to a nonmagnetic insulator takes place. 
The charge densities, as shown in Fig.~\ref{fig9}(a), 
 become disproportionated, mainly among the $b$-fMO sites. 
The disproportionation is coupled with the large difference 
 in $u_{\rm A}^{bb}$ among the two dimers [Fig.~\ref{fig9}(b)]. 
Therefore, similarly to the case of the AF states, 
 this is consistent with the discussions in Sect.~\ref{sec2} that 
 the singly occupied electrons 
 are mainly carried by the $b$-fMO. 
The total electron number for the four fMOs in each dimer 
 is also plotted in the upper panel of Fig.~\ref{fig9}(a). 
At large $1/K$, it approaches 6 and 4 for dimers X and Y, respectively. 
Since neutral dimers possess 4 electrons, 
 the electrons show a ``2-0"-type charge ordering. 

We consider that the mechanism for the instability toward the CLO state 
 is essentially analogous to that proposed in Ref.~\citen{Tamura_2004CPL}. 
The authors discussed the possible realization of a peculiar mechanism 
 based on the HOMO-LUMO inversion picture: 
 by distorting the intradimer $t_{ij}$ and then inducing a charge transfer between the dimers, 
 the electronic system gains energy. 
There are important distinctions between our calculation and that in Ref.~\citen{Tamura_2004CPL}: 
 one is that we included interdimer effects, 
 and also, the HOMO-LUMO mixing is considerable in our case, as we have discussed. 
However, the essential driving force for the occurrence of the CLO state remains the same, 
 i.e., 
 the lattice distortion realizing two types of dimer couples with the charge modulation, 
 leading to the insulating state due to the ``2-0" charge order. 
Therefore, it is a Peierls instability in a broad sense, 
 and we have confirmed that both $K_{\rm H}$ and $K_{\rm P}$ make CLO stable, 
 and they cooperate with each other. 
Nevertheless, to make this type of Peierls state stable in a strongly dimerized system,  
 the multiorbital nature is crucial. In single-orbital dimerized systems, 
 such ``2-0"-type charge order is hardly realized 
 in realistic parameter ranges. 

\subsection{Phase diagrams}\label{subsec33}
\begin{figure}
 \vspace*{1em}
 \begin{center}
  \includegraphics[width=0.5\textwidth]{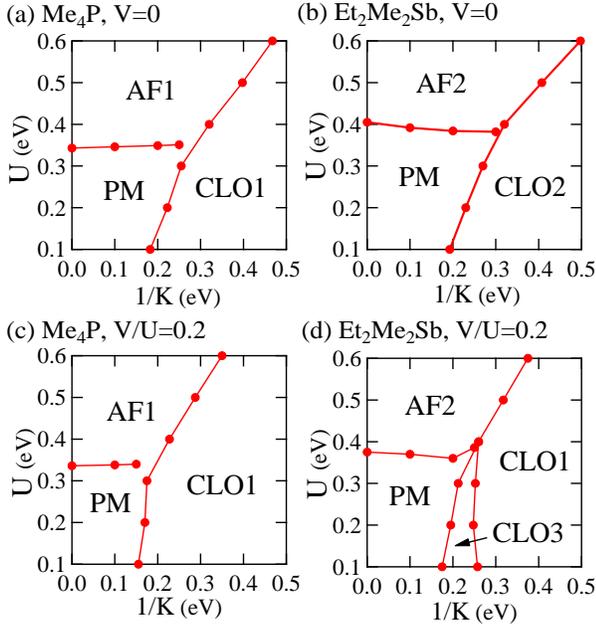}
 \end{center}
 \vspace*{-1em}
 \caption{%
 (Color online) 
Mean-field phase diagrams of the fragment model with varying on-site Coulomb energy $U$ and 
 electron-lattice couplings fixed as $K_{\rm H} = 4K_{\rm P} \equiv K$.
The parameters are for the cases of $V=0$ [(a) and (b)] and a fixed ratio of $V/U=0$ [(c) and (d)] 
 for $X=$ Me$_4$P [(a) and (c)] and Et$_2$Me$_2$Sb [(b) and (d)]. 
PM stands for the paramagnetic metallic phase while the AF and CLO phases 
 refer to those shown in Figs.~\ref{fig6} and \ref{fig8}, respectively. 
}\label{fig10}
 \vspace*{-1em}
\end{figure}

As we have seen in Sect.~\ref{subsec31}, 
 the stable AF pattern depends on the choice of $X$. 
Likewise, 
 the most stable CLO pattern among the self-consistent 
 solutions changes as the parameters are varied. 
In Fig.~\ref{fig10}, MF phase diagrams are shown, 
 where the electron correlation and electron-lattice coupling are varied. 
The former is controlled in the vertical axis $U$, 
 for $V=0$ [(a) and (b)] and the fixed ratio $V/U=0.2$ [(c) and (d)], 
 while the horizontal axis is $1/K$.  
The ratios among $V_{ij}$ are determined for each compound as mentioned above.  
We investigated the two sets of parameters obtained in Sect.~\ref{sec2} 
 for $X$ = Me$_4$P and Et$_2$Me$_2$Sb, i.e., the two end members. 

A common feature among the phase diagrams is that 
 there is competition among the paramagnetic metallic phase, 
 the AF insulating phase, and CLO insulating phases. 
All the phase boundaries are of first order. 
The AF insulator is favored in the large-$U$, small-$1/K$ region, as described in Sect.~\ref{subsec31}, 
 whereas, on the other hand, the CLO insulator is realized 
 in the small-$U$, large-$1/K$ region; 
 the electron-lattice coupling stabilizes the CLO state. 
Another point is that when $V_{ij}$ are included, 
 the CLO phases are widened; 
 therefore, the CLO state is also favored by the intersite Coulomb interaction. 

The difference between the two parameter sets for different $X$ 
 is not apparent concerning the phase boundaries, 
 but the stabilized AF and CLO patterns are different. 
The stable AF pattern is AF1 (AF2) for 
 $X$ = Me$_4$P (Et$_2$Me$_2$Sb), as seen in Sect.~\ref{subsec31} for the simple Hubbard model; 
 this does not change when ${\cal H}_{\rm P}, {\cal H}_{\rm H}$, and ${\cal H}_V$ are included. 
The stable CLO pattern for $V=0$ is 
 CLO1 (CLO2) for 
 $X$ = Me$_4$P (Et$_2$Me$_2$Sb). 
In fact, CLO1 (CLO2) has the same pattern as AF1 (AF2) 
 if one replaces dimers X and Y with up and down spins, respectively. 
This implies that the stability of the CLO state is related 
 to the degree of nesting of the Fermi surface 
 from the analogy to the AF state.  
However, for $V/U=0.2$, 
 while the stable pattern does not change for $X$ = Me$_4$P, 
 it sensitively changes for $X$ = Et$_2$Me$_2$Sb. 
As shown in Fig.~\ref{fig10}(d), 
 in the large-$1/K$ region, the stable pattern is now CLO1, 
 while there is a narrow region of the CLO3 phase near the phase boundary between the paramagnetic metal. 
Note that the calculation of Madelung energies in Ref.~\citen{Tamura_2005SM} 
 indicated similar values for patterns corresponding to CLO1, CLO3, and CLO4, 
 and a higher energy for CLO2; 
 this is consistent with our calculations including $V_{ij}$, 
 where CLO2 becomes unstable for $X=$ Et$_2$Me$_2$Sb. 
One can claim that there is a frustration effect also in the CLO states 
 owing to the triangular lattice structure of Pd(dmit)$_2$ dimers. 


\section{Discussion}\label{sec4}

In this section, we compare our results with previous works. 
First, let us comment on other theoretical works 
 taking into account the multiorbital nature of these materials. 
As metioned in Sect.~\ref{sec2}, 
 one important advantage of using the fMO scheme is that 
 an intuitive description based on the tight-binding picture 
 works well,  
 especially for the Coulomb interaction parameters~\cite{Tsuchiizu_2011JPSJ,Tsuchiizu_2012JCP}. 
In turn, when one transforms the basis sets from the fMOs to the original molecular orbitals 
 and then studies the model in terms of the latter, 
 the possibility of a counterintuitive choice of parameters should be taken into account. 
For example, 
 the Hund's rule coupling $J_\textrm{H}$ often becomes comparable to the on-site Coulomb terms. 
Complicated intermolecular four-body terms can appear and be nonnegligible. 
Such effects are not taken into account in previous works 
 based on the HOMO and LUMO of Pd(dmit)$_2$~\cite{Tamura_2004CPL,Mori_2000MCLC}. 
On the other hand, an alternative approach was chosen in Ref.~\citen{Nishioka_2013JPSJ} 
 to include intramolecular degrees of freedom, 
 whereas a microscopic footing for the basis set of the model was lacking. 

Next, the characteristic broken-symmetry states found in our calculations 
 owing to the multiorbital nature 
 are compared with experiments. 
Recently, an NMR experiment on the AF state in $X=$ Et$_2$Me$_2$P 
 has been performed~\cite{Otsuka_2014JPSJ}. 
The magnetic moment on each dimer was estimated to be $\simeq 0.28 \mu_\textrm{B}$, 
 which is considerably smaller than that obtained in our MF calculations. 
This should be a manifestation of quantum fluctuation owing to the spin frustration effect, 
 which is not included in our calculation, although the compound is located where 
 AF1 and AF2 patterns show similar MF energies. 
Moreover, 
 the asymmetry of the spin density on each Pd(dmit)$_2$ molecule 
 was pointed out~\cite{Otsuka_2014JPSJ}. 
This is in accordance with the difference in the magnetic moments on $a$- and $b$-fMOs 
 seen in our MF calculation (Fig.~\ref{fig7}). 
Nevertheless, 
 how to detect the existence of intramolecular antiparallel moments 
 as well as the 2D pattern AF1 
 predicted in the calculation remains as a future problem. 

We consider that the CLO state that we have obtained corresponds to that observed 
 in $X=$ Et$_2$Me$_2$Sb~\cite{Nakao_2005JPSJ,Tamura_2005CPL}, 
 while our calculation included assumptions, and 
 the electron-lattice coupling parameters 
 are not derived qualitatively. 
The X-ray crystal structure analysis data~\cite{Nakao_2005JPSJ}
 show that the C$=$C bonds on the two sides of the Pd(dmit)$_2$ molecule, 
 which provide information on the valence state, 
 become inequivalent in the low-temperature phase. 
Their lengths are 1.380 and 1.395 {\AA} at room temperature, 
 corresponding to the ligand where $a$- and $b$-fMOs are situated, respectively, 
 and split to 1.393 and 1.370 {\AA} (dimer X) and 
 1.394 and 1.412 {\AA} (dimer Y) at 10 K. 
In our MF calculation, the charge disproportionation occurs mainly on the $b$-fMO sites, 
 which is consistent with the data showing that a larger difference in the C$=$C bond length is observed there. 

Regarding the pattern of the CLO state, 
 although we found a narrow region of the CLO3 phase [Fig.~\ref{fig10}(d)], 
 consistent with the experiments, 
 we consider that additional factors can be important such as the interlayer couplings. 
In our calculations, only 2D models are considered, 
 and we find keen competition among different patterns depending on the parameters, 
 i.e., for different $X$ and strengths of interactions. 
Such competition was also found in the comparison of Madelung energies~\cite{Tamura_2005SM}. 
One point we note is that the CLO3 pattern in fact matches the solid crossing structure 
 in the $\beta'$-type Pd(dmit)$_2$ salts, and the system can gain the intersite Coulomb energy  
 by stacking the CLO3 pattern. 
We also note that, in another isostructural salt, Cs[Pd(dmit)$_2$]$_2$, 
 a metal-insulator transition is accompanied by a structural change~\cite{Underhill_1991JPCM}, 
 identified as having the same low-temperature structure 
 as that in $X=$ Et$_2$Me$_2$Sb~\cite{Nakao_2005JPSJ}. 
Although we have not analyzed the parameters for this salt, 
 we can deduce that the system is located near the phase boundary between 
 the paramagnetic metal and the CLO state in our phase diagrams in Fig.~\ref{fig10}. 
In fact, extended H\"{u}ckel calculations show larger overlap integrals than the series studied in this paper, 
 suggesting a larger bandwidth, which is consistent with such a closeness to the metallic state. 

One important question regarding this system is to find the controlling factor(s) 
 determining the ground states. 
In the experimental phase diagram, 
 the ground state varies as 
 AF $\rightarrow$ spin liquid $\rightarrow$ CLO. 
The revised parameters for the dimer model (Fig.~\ref{fig5}), 
 as pointed out in our previous study~\cite{Tsumuraya_2013JPSJ}, 
 imply that the degree of frustration is controlled by competing interchain interactions, 
 $T_\textrm{S}$ and $T_\textrm{r}$. 
The spin frustration is naively expected to be strongest when $T_\textrm{S} \simeq T_\textrm{r}$, 
 which is semiquantitatively consistent with 
 the destabilization of the AF phase when $X$ is varied from left to right. 
However, 
 the stability of the CLO phase for $X=$ Et$_2$Me$_2$Sb 
 remains elusive 
 since no significant difference in the MF phase diagram (Fig.~\ref{fig10}) 
 for the two material parameters is seen. 
One possibility is that the electron-lattice couplings become stronger 
 for $X$ on the right side of the experimental phase diagram. 
This is consistent with the fact that the unit cell volume is larger for 
 the compounds on the right, owing to the larger cation sizes. 
Then, softer electron-lattice couplings are expected 
 since there is more space for the Pd(dmit)$_2$ molecules
 to be displaced and deformed. 

\section{Summary}\label{sec5}

We have theoretically studied the electronic properties of $\beta'$-$X$[Pd(dmit)$_2$]$_2$ 
 using an effective model where fMOs serve as a basis set, 
 which is a suitable way of modeling multi-molecular orbital systems. 
The tight-binding parameters are estimated for different $X$ by fitting to first-principles band structures, 
 and the interaction effects are considered to discuss the broken-symmetry states. 
In contrast with previous discussions, 
 HOMO-LUMO mixing is appreciable and then disproportionation within each molecule is expected 
 in terms of charge density as well as spin density. 
This implies the need for revision of the picture of the effective dimer model 
 where the basis wave function constructing it is disproportionated. 
Broken-symmetry states are also affected by this mixing effect: 
 antiferromagnetic states show intramolecular antiparallel spin moments, 
 and charge-lattice coupled ordered states are mainly governed by one of 
 the fMOs. 
We have obtained mean-field phase diagrams for the competing ground states, 
 where both antiferromagnetic and charge-lattice orders can have different patterns within the 
 2D plane, depending on $X$ and the interaction parameters.  

\section*{Acknowledgment}
We thank T. Ito, K. Otsuka, and H. Sawa for discussions. 
This work was supported by Grants-in-Aid for Scientific Research
(Nos. 24108511, 26287070, and 26400377) 
from MEXT 
and the RIKEN iTHES project. 

\appendix


\section{%
$P$2$_1$/$m$ EtMe$_3$P salt}
\label{appB}

\begin{figure}[t]
 \begin{center}
  \includegraphics[width=0.4\textwidth,clip]{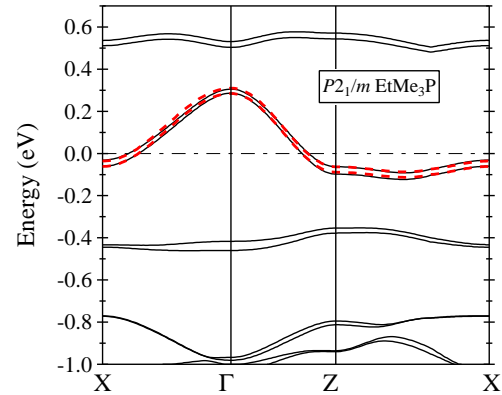}
 \end{center}
\setlength\abovecaptionskip{0pt}
 \caption{%
 (Color online) 
Electronic band structures near $\ef$ for the $P$2$_1$/$m$ phase of EtMe$_3$P[Pd(dmit)$_2$]$_2$. 
Solid and broken lines are the first-principles and fitted tight-binding bands, respectively. 
The fitting is for the effective dimer model, whose intralayer parameters are presented in Fig.~\ref{fig5}. 
 }\label{figB1} 
\end{figure}

We present the results of first-principles calculation of the electronic structure of 
 EtMe$_3$P[Pd(dmit)$_2$]$_2$ with
$P$2$_1$/$m$ symmetry~\cite{Kato_2006JACS}.  
Experimentally, this compound shows a phase transition to a lattice-distorted nonmagnetic state 
 called a valence bond order~\cite{Tamura_2006JPSJ}. 
The structure contains the same 2D stacking of Pd(dmit)$_2$ molecules 
 but a different interlayer stacking pattern. 
In Fig.~\ref{figB1}, 
 the first-principles band structure is shown, 
 where the intralayer dispersion is in fact overall similar to those of the $\beta'$-type compounds. 
The transfer integrals obtained by the dimer model fitting, shown in Fig.~\ref{fig5}, 
 are $T_\textrm{B} < T_\textrm{S} \simeq T_\textrm{r}$; 
 this is different from the other $\beta'$-type salts.

There is another difference from the $\beta'$-type salts discussed in the main text: 
 the interlayer parameters are larger. 
In the fitting based on the fMO scheme, 
 we needed two types of intermolecular transfer integrals in the interlayer directions
 for a reasonable fit, which are estimated as $-11$ and $-13$~meV. 
These are much larger than those for the $\beta'$-type salts on the order of 1 meV.

\end{document}